\documentclass[12pt]{article}
\usepackage{epsfig,psfig,graphicx,rotating}
\newcommand{\be}{\begin{equation}}
\newcommand{\ee}{\end{equation}}
\newcommand{\bea}{\begin{eqnarray}}
\newcommand{\eea}{\end{eqnarray}}
\begin{document}
\title{\bf UHECR as Decay Products of Heavy Relics? The Lifetime Problem}
\author{{\bf  H. J. de Vega$^{(a)}$ and
  N. S\'anchez$^{(b)}$}\\ \\
(a)  Laboratoire de Physique Th\'eorique et Hautes Energies, \\
Universit\'e Paris VI, Tour 16, 1er \'etage, 4, Place Jussieu, \\
75252 Paris, cedex 05, FRANCE. Laboratoire Associ\'e au CNRS UMR 7589.\\
(b) Observatoire de Paris,  LERMA, \\ 61, Avenue de l'Observatoire,
75014 Paris,  FRANCE. \\
Laboratoire Associ\'e au CNRS UMR 8540.}
\date{\today}
\maketitle
\begin{abstract}
The essential features underlying the top-down scenarii for UHECR are
discussed, namely, the stability (or lifetime) imposed to the heavy objects
(particles) whatever they be: topological and non-topological solitons,
X-particles, cosmic defects, microscopic black-holes, fundamental
strings. We provide an unified formula for the quantum decay rate
of all these objects as well as the particle decays in the standard
model. The key point in the top-down scenarii is the 
necessity to adjust the lifetime of the heavy object to the age of the
universe. This ad-hoc requirement needs a very high dimensional
operator to govern its decay and/or an extremely small coupling
constant. The natural lifetimes of such heavy objects are, however,
microscopic times associated to the GUT energy scale ($\sim
10^{-28}$sec. or shorter). It is at this energy scale (by the end of inflation)
where they could have been abundantly formed in the early universe and
it seems natural that they decayed shortly after being formed. 
\end{abstract}
\tableofcontents
\section{Introduction}

Ultrahigh Energy Cosmic Rays (UHECR) have been observed by a number of
experiments at energies above $10^{20}$ eV\cite{exp}. 
The forthcoming cosmic rays detectors as the Auger array, the EUSO and
OWL space observatories are expected to greatly improve our present knowledge
on the UHECR gathered from Fly's Eye, HiRes, AGASA and previous
detectors\cite{exp,euso}.  

Top-down scenarii for UHECR are based on heavy relics from the early
universe which are assumed to decay at the present time. In all cases,
whatever the nature of 
the objects: heavy particles, topological and non-topological solitons,
black-holes, microscopic fundamental strings, cosmic defects etc., one
has to fine tune the lifetime of these objects to be the age of the universe.

We provide an unified description for the quantum decay formula of
unstable particles which encompass all the above mentioned cases,
as well as the particle decays in the standard model (muons, Higgs,
etc). In all cases the decay rate can be written as,
\be \label{formae}
\Gamma = \frac{g^2 \; m }{\mbox{numerical factor}}
\ee
where $g$ is the coupling constant, $m$ is the typical mass in the theory
(it could be the mass of the unstable particle) and the numerical
factor contains often relevant mass ratios for the decay process.

The key drawback of all top-down scenarii is the lifetime problem. The
ad-hoc requirement of a lifetime of the order the age of the universe
for the heavy particles implies an operator with a very high dimension
describing the decay, and/or an extremely small coupling constant.

Heavy relics could have been formed by the end of inflation at typical
GUT's energy scales, but their natural lifetime would be of the order
of microscopic times typically associated to GUT's energy scales
\cite{resp,big}. 

UHECR may result from the acceleration of protons and ions by
shock-waves in astrophysical plasmas (Fermi acceleration
mechanism)\cite{raycos}. Large enough sources can accelerate particles to the
energies of the observed UHECR. Sources in the vicinity of our galaxy
as hot spots of radio galaxies (working surfaces of jets and the inter
galactic medium) and blazars (active galactic nuclei with relativistic
jet directed along the line of sight) as BL Lacertae can evade the GZK
bound\cite{stec,domo,raycos2}. 

\section{Topological solitons, non-topological solitons and heavy particles}
Stable  solutions in classical field theory (as monopoles)
become (heavy) particles in quantum  field theory. There is no difference
at the quantum level between heavy particles associated to a local
field and those associated to classical stable solutions. 

The stability of classical solutions in field theory is a highly
nontrivial issue. There are basically two types of solutions:
topological and non-topological. Topological  classical solutions have
associated a non-zero topological number (topological charge)
which vanishes for the
vacuum. If there is a lower bound for the energy of the solution
involving this topological number  the classical solution is
stable. This is the case for kinks in one space dimension scalar theories, 
vortices in the two-dimensional Higgs model\cite{bogo}, monopoles in the three
dimensional Georgi-Glashow model\cite{thpo}, Hopf solitons in
appropriate  three dimensional scalar models\cite{hopf}.  In all known cases,
classical stability comes together with quantum stability.

Gravitational analogues of these classical solutions exist in the
euclidean (imaginary time) regime \cite{gh,ns}: they are black-holes
in three space dimensions (with periodicity in the imaginary time),
which are gravitational analogues of electric type monopoles and
Taub-Nut's in four space dimensions (gravitational analogues of
magnetic type monopoles). The topological charges here are related to
the temperature and magnetic charge of the solutions,
respectively\cite{ns,mg}.  

It must be stressed that the mere presence of a
conserved topological number does not guarantee the stability of the
corresponding classical solution. The  energy  must be
related with the topological number in question such that a non-zero
topological 
number implies a non-zero energy\cite{colem}. Otherwise, a classical solution
possessing non-zero topological number can decay into lighter particles.

In other words, the topological charge may be disconnected from the
dynamics and it can decay in the course of the evolution. A
topological soliton may collapse loosing its topological charge.  
This does not happen when the topological charge bounds the energy
from below. 

Non-topological solitons are stable thanks to a conserved $ U(1) $
charge of `electric' type\cite{td}. Again, the mere presence of a
conserved $ U(1) $ 
charge does not guarantee stability for charged particles except for
the lightest one. Let us call $m$ the mass of the  lightest charged
particle and let us take its $ U(1) $ charge as unit of charge. 
Assume that there are heavier particles with mass $ M > m $ and charge
$ Q > 1 $ with $ M = M(Q) $. A sufficient condition for quantum
stability is 
$$
 M(Q) < m \; Q \; ,
$$
since a particle with charge $ Q $ and mass larger or equal than $  m
\; Q $ can always decay into $ Q $ particles of mass $ m $ and unit charge
respecting charge and energy conservation.

It must be stressed that in quantum theory all non-forbidden process
{\bf do happen}.

\section{Quantum Decay of Heavy Particles}

Typically, the decay of a heavy particle with mass $ m_X $ can be
described by an effective interaction lagrangian formed by the local
field $ X(x) $ associated to this  heavy particle times the lighter
fields in which it decays. Let us take the muon decay which is a well
known case. Notice the mass of the muon $ m_{\mu} = 206.8 \, m_e
\gg m_e $.

The effective Fermi lagrangian can be written as\cite{libro}
\be\label{effL}
{\cal L}_I =  -{G_F \over \sqrt2 } \; {\bar \psi}_{\nu_\mu}
\gamma^{\alpha}(1 + \gamma_5) \psi_{\mu} \; {\bar
\psi}_{\nu_e}\gamma_{\alpha}(1 + \gamma_5) \psi_{\nu_e}
\ee
where $ \psi_{\mu} $ stands for the muon field, $ \psi_{\nu_e} $ and $
{\psi}_{\nu_\mu} $ for the electron neutrino and muon neutrino fields,
respectively. The Fermi coupling $ G_F $ has the dimension of an inverse square
mass.  

The muon width $ \Gamma_{\mu} $ describing the decay is then given by
$$
\Gamma_{\mu} = { G_F^2 \; m_{\mu}^5 \over 192 \, \pi^3 }
$$
The Fermi coupling can be related to the W-mass as follows
$$
{G_F \over \sqrt2 }= {g^2 \over 8 \; m_W^2}
$$
where $g$, the standard model coupling, is dimensionless. Thus,
\be \label{mudecay}
\Gamma_{\mu} = { g^4 \; m_{\mu} \over 6144 \, \pi^3 } \left({m_{\mu} \over
m_W}\right)^4
\ee
As we shall see below, Eq.(\ref{mudecay}) has the generic structure
for the decay width of an unstable particle.

For the muon decay, the monomial interaction in the effective
lagrangian (\ref{effL}) has dimension six in mass units. 

An analogous example is the Higgs decay into muons, neutrinos,
$W^{\pm}$ and the $Z^0$. Notice that the Higgs mass $ m_H $ must
be higher than the  $W^{\pm}$ mass $m_W$ and the $Z^0$
mass. The  lagrangian as given by the standard model is here
\be\label{effLH}
2 \, g \sin \theta_W \; M_W   \; H \; W^+_{\mu}  \; W_-^{\mu}
\ee
and a similar expression for the coupling with the $ Z $. Here $
\theta_W $ stands for Weinberg's angle.

One finds for the Higgs decay rate\cite{tito},
\be \label{higdeca}
\Gamma_{Higgs} = {3 \, g^2 \over 128 \, \pi} \; m_H \left({m_{H} \over
m_W}\right)^2
\ee
where we consider for simplicity the case $ M_H \gg M_W $. 
In this case the  monomial interaction in the effective
lagrangian (\ref{effLH}) has dimension three in mass units. 

Notice that in both cases, eq.(\ref{mudecay}) and eq.(\ref{higdeca}), the
width grows as a positive power of the mass of the decaying particle.

Let us consider an effective lagrangian containing a local monomial of
dimension $n$ (in mass units) 
\be \label{lefX}
{\cal L}_I = {g \over M^{n-4}} \, X \Theta \; .
\ee
Here the field $ X $ is associated to the decaying particle of mass $
m_X $ and $ \Theta $ stands for the product of fields coupled to it.

Then, the  decay rate for a particle of mass $ m_X $ takes the form 
\be \label{anchX}
\Gamma = { g^2 \over \mbox{numerical factor}}  \; m_X \; \left( {m_X
\over M}\right)^{|2n-8|}
\ee
$\Gamma_{\mu}$ eq.(\ref{mudecay}) and $\Gamma_{Higgs}$
eq.(\ref{higdeca}) correspond to $n=6$ and $n=3$, respectively. 

\section{Quantum Decay of Solitons}

The mass of classical soliton solutions (as magnetic monopoles in
unified theories)  are of the form
$$
M_{sol} = {\mu \over g^2}
$$
where $ \mu $ is the mass of the basic fields in the lagrangian and $
g $ their dimensionless coupling. For small coupling these objects are
much heavier than the particles associated to the basic fields in the
lagrangian. 

Quantum mechanically the soliton mass acquires corrections of order $
g^0 $ and higher. To one-loop level one finds
\be \label{ptocero}
M_{sol} = {\mu \over g^2} + \frac12 \sum_n \left[ \omega_n - \omega_n^0
\right] \; ,
\ee
where $ \omega_n $ stands for the frequency of oscillations around
the soliton. These oscillations are close but not identical to the
frequency  of oscillations around the vacuum $ \omega_n^0 $. The sum
in eq.(\ref{ptocero}) yields a finite result proportional to $ \mu
$\cite{colem}. 

Now, if the classical solution is unstable some of the frequencies $
\omega_n $ develops an imaginary part $ i \mu \beta $ where $ \beta $
is a pure number. 
Hence,
\be \label{insol}
Im M_{sol} = \beta \; \mu\quad \mbox{and} \quad Re  M_{sol} = {\mu \over g^2} +
{\cal O} (g^0)
\ee
and
\be \label{anchoS}
\Gamma_{sol} = Im M_{sol} = g^2 \; \beta\; Re  M_{sol}
\ee
We see that the width $\Gamma_{sol}$ has a similar structure than for heavy
particles in the previous section.

The term $ {\cal O} (g^0) $ in eq.(\ref{insol}) stand for the first quantum
correction to the mass. Notice that we choose $ \hbar = 1 $ which is
absorbed in $ g^2 $. 

\section{Quantum Decay of Fundamental Strings}

The decay of closed strings in string theory has been computed to the
dominant order (one string loop)\cite{cuer}.
Assuming the closed string in an $N$th excited state, it can
decay into lower excited states including the graviton and the
dilaton. The mass of this quantum string is given by
$$
m^2 = 32\pi \; T \; N
$$
where $ T $ is the string tension $ T = 1/(4\pi \alpha') $ and
$\alpha'$ the string constant. The
length of such string is given by $ L = 2\; \alpha' \; m $. 

One finds for the total width for string decay\cite{cuer},
\be \label{ancue}
\Gamma_{string} ={ \kappa^2 \; \sqrt{T \; N} \over { \mbox{numerical
factor}}} 
\ee
where the dimensionless coupling $\kappa$ is given by,
$$ 
\kappa = 48\pi \sqrt{2 G T}
$$
The total width can be then rewritten as,
\be \label{ancue2}
\Gamma_{string} = { \kappa^2 \; m \over 1083 \times \mbox{numerical
factor}} \; .
\ee
This formula again has the same structure as the previous widths
(\ref{anchX}) and (\ref{anchoS}) once we identify $ g =
\kappa , \;  m_X = Re M_S = m $.

Eq.(\ref{ancue2}) can be rewritten as,
\be \label{ancue3}
\Gamma_{string} = 42 \; \frac{G \, T  \, m}{\mbox{numerical factor}}
= \frac{21}{16 \, \pi} \; \frac{G \;
m^3}{\mbox{numerical factor}} = \frac{21\sqrt{2}}{\mbox{numerical
factor}} \; \frac{\sqrt{N} \; G }{\alpha'^{3/2}} 
\ee

\section{Quantum Decay of Black Holes}

As it is known, in the context of field theory black holes decay
semiclassically through thermal emission at the Hawking temperature\cite{sw}
$$
T_{BH} = \frac{\hbar \, c}{4 \, \pi \, k_B} \; \frac1{R_s} \quad ,
\quad R_s = \frac{2 \, G \, M}{c^2}
$$
($M$ being the black hole mass and $k_B$ the Boltzmann constant).

Black hole emission follows a `gray body' spectrum (the `filter' being
the black hole absorption cross section $\sim R_s^2 $). The mass loss
rate in this process can be estimated following a Stefan-Boltzmann
relation,
$$
\frac{dM}{dt} = - \sigma \; R_s^2 \; T_{BH}^4 \sim T_{BH}^2
$$
where $ \sigma $ is a constant. Thus, the black hole decay rate is
$$
\Gamma_{BH} = \left| \frac1{M} \; \frac{dM}{dt} \right| \sim G \;
T_{BH}^3 \sim \frac{G}{R_s^3}
$$
As evaporation proceeds, the black hole temperature increases till it
reaches the string temperature\cite{marnor}
$$
T_{string} =\frac{\hbar \, c}{k_B} \; \frac1{b \; L_s} \quad ,
\quad L_s = \sqrt{\frac{ \hbar \, \alpha'}{c}}
$$
($L_s$ being the fundamental string length and $b$ a constant
exclusively depending on the spacetime dimensionality and the string
model chosen.) The black hole enters its string regime
$ T_{BH} \to T_{string}, \; R_s \to L_s $, becomes a string state and
decays with a width
$$
\Gamma_{BH} \to G \; T_s^3 \sim \frac{G}{ \alpha'^{3/2}} \sim
\Gamma_{string} \; . 
$$
Notice that this formula is similar to eq.(\ref{ancue3}) and again has
the generic structure of the widths eq.(\ref{anchX})-(\ref{anchoS})
and (\ref{ancue2}) if one identifies  $ g = \kappa , \;  m_X = Re M_S = m $.

We consider here both fundamental strings and black holes since their
decay rates can be nicely recasted as in eq.(\ref{formae}) independently of
whether or not they may be considered as candidate sources of UHECR.

\section{Particles Lifetime and the Age of the Universe}

Heavy particles with masses in the GUT scale can be produced in large
numbers during inflation and just after inflation\cite{resp}. The production
mechanism is parametric or spinodal amplification in the inflaton
field. That is, 
linear resonance of the quantum modes of the heavy field in the
background or condensate of the inflaton. In addition, non-linear quantum
phenomena play a crucial role and can enhance the particle
production\cite{big}. Such non-linear production is of the same order
of magnitude as the gravitational production of particles by the time
dependent metric. 

Once these heavy particles are produced they must have a lifetime of
the order of the age of the universe in order to survive in the
present universe and decay into UHE cosmic rays. Only in the early
universe the production of such heavy objects is feasible due to their
large mass.  

Moreover, in order to be the source of  UHECR, these
particles must have a mass of the observed UHECR, namely $ m_X > 
10^{21}$ eV = $ 10^{12}$ GeV. 

Let us assume that the effective lagrangian (\ref{lefX}) describes the
decay of the $ X $ particles\cite{ell}. Their lifetime will be given
by eq.(\ref{anchX}) 
$$
\tau_X = { \mbox{numerical factor}\over g^2 } \; { 1 \over m_X} \;
\left( {M \over m_X}\right)^{2n-8} =  { \mbox{numerical factor}\over
g^2 } \; { 1 \over m_X} \; 10^{6(n-4)} 
$$
where we set a GUT mass $ M = 10^{15}$ GeV. The age of the
universe is $ \tau_{universe} \sim 2 \; 10^{10}$years and we have to
require that $ \tau_X > 
\tau_{universe} $. Therefore,
\be \label{taunive}
10^{54} < 
{ \mbox{numerical factor}\over g^2 } \; 10^{6(n-4)}
\quad \mbox{or}  \quad  \log_{10} g < 
3 (n - 13)
\ee
and we dropped the numerical factor in the last step.

For $ g \sim 1 $, eq.(\ref{taunive}) requires an operator $
\Theta $ with dimension at least thirteen in the effective lagrangian
(\ref{lefX}) which is a pretty high dimension. 

That is, one needs to exclude all operators of dimension lower than
thirteen in order to 
extremely suppress the decay. Clearly, one may accept lower dimension operators
$ \Theta $ paying the price of a small coupling $ g $. For example: $
g = 10^{-9} $ and $ n = 10 $ fullfil the above bound still being a
pretty high dimension operator.  Notice that a moderate $n$ as $n=4$
lowers the coupling to $g \sim 10^{-27} $.

In summary, a heavy X-particle can {\bf survive} from the early
universe till the present times if one chooses

\begin{itemize}
\item{an extremely small coupling $g$}

$\quad  $             and/or

\item{an operator $\Theta$ with high enough dimension}
\end{itemize}

None of these assumptions can be supported by arguments other than
imposing a lifetime of the age of the universe to the X-particle. That
is, the lifetime must be here {\bf fine tunned}. That is, one has to
built an {\bf ad-hoc} lagrangian to describe the X-particle decay. 
Indeed, a variety of  ad-hoc lagrangians have been proposed in the
literature together with the symmetries which can {\bf adjust} a wide
variety of lifetimes\cite{benham}. 

It must be recalled that no known (weakly broken) symmetry protects
the X-particle from decaying rapidly, except for supersymmetry. 
However, if supersymmetry would be invoked in this context, that would
imply that the supersymmetry scale is at the GUT scale or beyond.
It must be also noticed that the {\bf natural lifetime} for particles of
such a mass is the GUT scale, that is typically $ 10^{-28}$ sec. - $
10^{-35}$ sec. 

\section{Cosmic Defects and Heavy Particles}

Closed vortices from abelian and non-abelian gauge theories are {\bf not}
topologically stable in $3+1$ space-time dimensions. Static vortices
in $3+1$ space-time dimensions just collapse to a point since their
energy is proportional to their length. They do that in a very short
(microscopic) time. 

It must be noticed that only a restricted set of spontaneously
broken non-abelian gauge theories exhibit vortex solutions. For
example, there are no topologically stable vortices in the standard
$ G = SU(3) \times SU(2)  \times U(1) $ model in $3+1$ space-time
dimensions just because $ \Pi_1(G) $ and $ \Pi_2(G) $  are trivial for
such group manifolds. [For a recent review see \cite{rmp}].
Grand unified theories may or may not 
posses vortex solutions in $2+1$ space-time dimensions
depending under which  representations of the
gauge group belong the Higgs fields.

The existence of cosmic string networks is not established although
they have been the subject of many works. In case such networks would
have existed in the early universe they may have produced heavy
particles X of the type discussed before and all the discussion on
their lifetime applies here. The discussion on the lifetime problem
also applies to rotating superconducting strings which  have been proposed as
classically stable objects\cite{vort}. 

Cosmic strings are closed vortices of horizon size. 
In $3+1$ space-time dimensions, strings collapse very fast
except if they have horizon size in which case their lifetime would be
of the order of the age of the universe. However, such horizon size
cosmic strings are excluded by the CMB anisotropy observations and by
the isotropy of cosmic rays. 
Such gigantic objects behaves classically whereas microscopic closed
strings (for energies $ < 
M_{Planck} = 10^{19}$ GeV) behave quantum mechanically. 

In summary, a key point here is the {\bf unstability} of topological defects in
$3+1$ space-time dimensions. Unless one chooses very specific models
\cite{thpo,hopf,td,gh,ns,mg}, topological defects decay {\bf even
classically} with a short lifetime. They collapse to a point at a
speed of the order of the speed of light in $3+1$ space-time dimensions. 

\section{Decay Products of Heavy Particles and Final Conclusions}

In summary, if the X-particles, whatever their origin and type could
be made {\bf sufficiently stable} to survive till now, then their
decay products could provide the UHECR observed today.
However, the X-particles lifetime of the order of the age of the
universe must be imposed ad-hoc i. e. fine tuned while the natural
lifetime for those particles should be extremely short about $
10^{-28}$ sec at most.

Various GUTs contain candidates for X-particles of masses around the
GUT scale ranging approximately from $10^{12}$ Gev to $10^{16}$ Gev
depending on the model. These particles could have been 
produced  naturally in the early universe typically by the end of
inflation\cite{resp}.  Analogously,  topological defects, fundamental
strings and primordial black holes could have been formed in the early
universe. The hard job, however,  is to have these heavy objects still present
and decaying today.  Instead of that,
it seems more natural that the X-particles and the other heavy objects
above mentioned decayed in the early universe shortly after being
formed, having lifetimes corresponding to their respective energy
scales. Their decay products will then form relic primordial
backgrounds as graviton, neutrino and dilaton backgrounds, as we know now 
the relic photon CMB background. Those backgrounds could
have characteristic detectable spectra and signatures containing informations
about the early universe.


\begin{thebibliography}{99} 


\bibitem{exp} J. Linsley, Phys. Rev. Lett., 10, 146 (1963).

M. A. Lawrence et al. J. Phys. G 17, 773 (1991).

D. J. Bird et al. Ap J, 441, 144 (1995).

D. Kieda et al. Proc. 26th ICRC, Salt Lake City, Utah, USA.

M. Teshima, Proceedings of TAUP 2001, L'Aquila, Sep. 2001.

AGASA: http://www-akeno.icrr.u-tokyo.ac.jp/AGASA/

\bibitem{euso} L. Scarsi, 
Lectures given at the D. Chalonge School in
Astrofundamental Physics, in `The Cosmic Microwave Background', 
ed.  N. G. S\'anchez (Dordrecht: Kluwer) p. 483 (2000).

EUSO: http://www.ifcai.pa.cnr.it/~EUSO

\bibitem{raycos} R. Blanford, D.  Eichler, Phys. Rep. 154, 1 (1987).

L O'C Drury, Rep. Prog. Phys. 46, 973 (1983). 

Y. A. Gallant, A. Achterberg, MNRAS, 305, L6 (1998).
A. Achterberg, Y. A. Gallant, J. G. Kirk, A. W. Guthmann, astro-ph/0107530.

F. Halzen, astro-ph/0111059

M. A. Malkov and P. H. Diamond, astro-ph/0102373,
M. A. Malkov and P. H. Diamond, H. J. V\"olk, ApJ, 533, L171 (2000). 

P. Blasi, astro-ph/0110401.

T. W. Jones, astro-ph/0012483.

M. Ostrowski, astro-ph/0101053.

M. Vietri, astro-ph/0002269.

\bibitem{stec} F. W. Stecker, Lectures given at the D. Chalonge School in
Astrofundamental Physics, in `Phase Transitions in the Early Universe:
Theory and Observations', ed. H. J. de Vega, I. M. Khalatnikov and
N. G. Sanchez (Dordrecht: Kluwer) p. 485 (2001) [astro-ph/0101072].

P. L. Biermann,  Lectures given at the D. Chalonge School in
Astrofundamental Physics, in `Phase Transitions in the Early Universe:
Theory and Observations', ed. H. J. de Vega, I. M. Khalatnikov and
N. G. Sanchez (Dordrecht: Kluwer) p. 505 (2001).

\bibitem{domo} G. Domokos, S. Kovesi-Domokos, hep-ph/0107095. 

\bibitem{raycos2} P. L. Biermann, P. Strittmatter,
Astopart. Phys. 322, 643 (1987). 
J. P. Rachen, P. L. Biermann, A \& A, 272, 161 (1993).

P. G. Tinyakov and I. I. Tkachev, astro-ph/0111305.

\bibitem{bogo} A. A.  Abrikosov, Zh. Eksp. Teor. Fiz. 32, 1442  (1957).

H. B. Nielsen and P. Olesen, Nucl. Phys. B61, 45 (1973).

E. B. Bogomolny, Sov. J. Nucl. Phys. 24, 449 (1976).

 H. J. de Vega and F.A. Schaposnik, Phys. Rev. D14, 1100 (1976). 

\bibitem{thpo} G. t' Hooft, Nucl. Phys. B79, 276 (1974).

A. M. Polyakov, JETP Lett. 20, 194 (1974).

\bibitem{hopf} H. J. de Vega, Phys. Rev. D18, 2945 (1978).

\bibitem{colem} See for example, S. Coleman, Erice Lectures 1975.

\bibitem{td} See for a review T. D. Lee and Y. Pang, Phys. Rep. 221,
251 (1992). 

\bibitem{gh} G. W. Gibbons and S. W. Hawking, Comm. Math. Phys. {\bf
66}, 291 (1978). 

\bibitem{ns} N. S\'anchez, `Einstein equations, Non-linear sigma
models and self-dual Yang-Mills theory', in `Differential Geometric
Methods in Mathematical Physics', Lecture Notes in Mathematics,
vol. 139, Springer Verlag (1985).

\bibitem{mg} N. S\'anchez, `Topological Invariants and Thermal
properties of analytic mappings', in Proceedings of the Marcel
Grossman Meeting, p. 501-518 (1982), North Holland.  

\bibitem{libro}  See for example, J. F. Donoghue et al., Dynamics of
the Standard Model, Cambridge Univ. Press, 1992. 

\bibitem{tito} B. A. Kniehl and A. Sirlin, Phys. Lett. B440, 136 (1998).

\bibitem{cuer} R. B. Wilkinson, N. Turok, D. Mitchell,
Nucl. Phys. B332, 131 (1990).

J. Dai, J. Polchinski, Phys. Lett. B220, 387 (1989). 

\bibitem{sw} S. W. Hawking, Comm. Math. Phys. {\bf 43}, 199 (1973).

\bibitem{marnor} M. Ram\'on Medrano, N. S\'anchez, Phys. Rev. {\bf
D61}, 084030 (2000). 

\bibitem{resp} D. J. H. Chung, E. W. Kolb and A. Riotto,
Phys. Rev. D60,  063504 (1999).

D. Boyanovsky, M. D'Attanasio,  H. J. de Vega, R. Holman and
D.-S. Lee,

Phys. Rev. {\bf D52}, 6805 (1995).  

\bibitem{big} D. Boyanovsky, C. Destri, H. J. de Vega, R. Holman and
J. F. J. Salgado, 

Phys. Rev. {\bf D57}, 7388 (1998).

\bibitem{ell} J. Ellis, Nuovo Cim. 24C, 483 (2001).
J. Ellis, J. L. Lopez and D. V. Nanopoulos, Phys. Lett. B247, 257
(1990). 

S. Sarkar, R. Toldra, Nucl. Phys. B621,495  (2002).

\bibitem{benham}  K. Hamaguchi, Y. Nomura, T. Yanagida, Phys. Rev. D58
(1998) 103503, D59 (1999) 063507.

K. Benakli, J. Ellis, D. V. Nanopoulos, Phys. Rev. D59 (1999) 047301.

K. Hamaguchi, Izawa K.-I., Y. Nomura, T. Yanagida, Phys. Rev. D60
(1999) 125009 .

\bibitem{vort} R. L. Davis, E. P. S. Shellard, Nucl. Phys. B323, 209 (1989). 

X. Martin , Phys. Rev. D51, 4092 (1995).

A. L. Larsen, M. Axenides, Class. Quant. Grav. 14, 443 (1997).

\bibitem{rmp} I. S. Aranson, L Kramer, Revs. Mod. Phys. 74, 99
(2002). 

\end{thebibliography}
\end{document}